\documentclass[pre,floats, superscriptaddress, longbibliography, dvipsnames,nofootinbib]{revtex4-2}
\usepackage[utf8]{inputenc}
\usepackage{amssymb}
\usepackage{amsmath}
\usepackage{tikz}
\usepackage{hyperref}

\usetikzlibrary{patterns}
\usetikzlibrary{arrows.meta,matrix,%
                    decorations.pathreplacing}
\newcounter{bla}

\newcommand{\bi}{\begin{itemize}}
\newcommand{\ei}{\end{itemize}}
\newcommand{\be}{\begin{enumerate}}
\newcommand{\ee}{\end{enumerate}}

\usetikzlibrary{patterns}
\usepackage{luatex85}
\newlength{\flexcheckerboardsize}

\newcommand{\defineflexcheckerboard}[4]{
    \setlength{\flexcheckerboardsize}{#2}
    \pgfdeclarepatterninherentlycolored{#1}
        {\pgfpointorigin}{\pgfqpoint{2\flexcheckerboardsize}    
        {2\flexcheckerboardsize}}
        {\pgfqpoint{2\flexcheckerboardsize}
        {2\flexcheckerboardsize}}%
        {
            \pgfsetfillcolor{#4}
            \pgfpathrectangle{\pgfpointorigin}{
            \pgfqpoint{2.1\flexcheckerboardsize}    
                {2.1\flexcheckerboardsize}}
          \pgfusepath{fill}
          \pgfsetfillcolor{#3}
          \pgfpathrectangle{\pgfpointorigin}
            {\pgfqpoint{\flexcheckerboardsize}
            {\flexcheckerboardsize}}
          \pgfpathrectangle{\pgfqpoint{\flexcheckerboardsize}
            {\flexcheckerboardsize}}
            {\pgfqpoint{\flexcheckerboardsize}
            {\flexcheckerboardsize}}
            \pgfusepath{fill}
        }
}

\defineflexcheckerboard{flexcheckerboard_redblue}{.5cm}{red}{blue}

\makeatletter
\newcommand*{\balancecolsandclearpage}{%
  \close@column@grid
  \cleardoublepage
}

\begin{document}

\title{The QISG suite: high-performance codes for studying Quantum Ising Spin Glasses}

\author{Massimo Bernaschi}\email{massimo.bernaschi@cnr.it}\affiliation{Istituto
  per le Applicazioni del Calcolo, CNR, Via dei Taurini 19, 00185 Rome, Italy}

\author{Isidoro González-Adalid Pemartín}\email{isidorog@ucm.es (Corresponding
author)}\affiliation{Departamento de
  F\'\i{}sica Te\'orica, Universidad Complutense, 28040 Madrid, Spain}

\author{Víctor Martín-Mayor}\email{vicmarti@ucm.es}\affiliation{Departamento de
  F\'\i{}sica Te\'orica, Universidad Complutense, 28040 Madrid, Spain}

\author{Giorgio Parisi}\email{giorgio.parisi@uniroma1.it}\affiliation{Dipartimento di Fisica, Sapienza
  Università di Roma, P.le Aldo Moro 2, 00185 Rome,
  Italy}\affiliation{Nanotec-Rome unit, CNR,  P.le Aldo Moro 2, 00185 Rome, Italy}
\date{\today}

\begin{abstract}
We release a set of GPU programs for the study of the Quantum ($S=1/2$)
Spin Glass on a square lattice, with binary couplings. The library contains
two main codes: {\bf MCQSG} (that carries out Monte Carlo simulations using
both the Metropolis and the Parallel Tempering algorithms, for the problem formulated in the Trotter-Suzuki
approximation), and {\bf EDQSG} (that obtains the extremal eigenvalues of the
Transfer Matrix using the Lanczos algorithm). EDQSG has allowed us to diagonalize
transfer matrices with size up to $2^{36}\times2^{36}$. From its side, MCQSG
running on four NVIDIA A100 cards delivers a sub-picosecond time per
spin-update, a performance that is competitive with dedicated hardware. We include
as well in our library GPU programs for the analysis of the spin
configurations generated by MCQSG. Finally, we provide two auxiliary codes: the first
generates the lookup tables employed by the random number generator of MCQSG; the second one
simplifies the execution of multiple runs using different input data.
\end{abstract}


\maketitle

\textbf{PROGRAM SUMMARY/NEW VERSION PROGRAM SUMMARY}\\

\begin{small}
\noindent
{\em Program Title: QISG Suite}                                          \\
{\em CPC Library link to program files:} \url{https://doi.org/10.17632/g97sn2t8z2.1} \\
{\em Licensing provisions: MIT }\\
{\em Programming language: CUDA-C} \\
{\em Nature of problem:} The critical properties of quantum disordered systems are known only
in a few, simple, cases whereas there is a growing interest in gaining a better understanding of their behaviour due to
the potential application of quantum annealing techniques for solving optimization problems. In this context,
we provide a suite of codes, that we have recently developed, to the purpose of studying the 2D Quantum Ising Spin Glass.\\
{\em Solution method:} We provide a highly tuned multi-GPU code for the Montecarlo simulation of the 2D QISG based on a combination of Metropolis and Parallel Tempering algorithms. Moreover, we provide a code for the evaluation of the eigenvalues of the transfer matrix of the 2D QISG for size up to L=6. The eigenvalues are computed by using the classic Lanczos algorithm that, however, relies on a custom multi-GPU-CPU matrix-vector product that speeds-up dramatically the execution of the algorithm.\\
   \\
\end{small}

\section{Introduction}
We release highly-tuned multi-GPU codes developed for exploring the
critical properties of a paradigmatic quantum disordered system: the
Ising spin glass in 2 spatial dimensions, evolving under the
Hamiltonian 
\begin{equation}\label{IQSG}
H=-\sum_{\langle \boldsymbol{x}, \boldsymbol{y}\rangle}J_{\boldsymbol{x},\boldsymbol{y}}\sigma^Z_{\boldsymbol{x}}\sigma^Z_{\boldsymbol{y}} -\varGamma\sum_{\boldsymbol{x}}\sigma^X_{\boldsymbol{x}}\,,
\end{equation}
that contains both spin couplings
$J_{\boldsymbol{x},\boldsymbol{y}}=\pm 1$ (that may induce
frustration), and a transverse field $\varGamma$. Specifically, the
quantum spin operators $\sigma^Z_{\boldsymbol{x}}$ and
$\sigma^X_{\boldsymbol{x}}$ are, respectively, the third ($Z$) and the
first ($X$) Pauli matrices, acting on the nodes $\boldsymbol{x}$ of a
square lattice endowed with periodic boundary conditions. We indicate with $L$
the linear size of the square lattice. The interaction term
$J_{\boldsymbol{x},\boldsymbol{y}}\sigma^Z_{\boldsymbol{x}}\sigma^Z_{\boldsymbol{y}}$
is restricted to lattice nearest-neighbors.

At very low temperatures, {\em quantum} fluctuations become more and
more important up to the point that, at zero temperature, as the ratio
between the coupling energy and the transverse field is varied, the
system \eqref{IQSG} undergoes a quantum phase transition where its
ground state properties abruptly change.

Beyond the scientific interest (the critical properties of a quantum
disordered system are known only in very few, and peculiar cases --
mainly one-dimensional geometries), a better understanding of the
behaviour of this system has important practical implications in
light of the second quantum revolution, where mesoscopic quantum
systems can unlock new and potentially game-changing technologies for
solving many real-world optimization problems.

Our work requires a preliminary reformulation of the problem based on the
Suzuki-Trotter formula \cite{ST}, such that the quantum system
\eqref{IQSG} in two dimensions can be studied as if it were a classic
system in $2+1$ dimensions. The quantum spin operator at site
$\boldsymbol{x}$ is replaced by a chain of classical variables
$S_{\boldsymbol{x},\tau}=\pm 1$ with $\tau=0,1,\ldots, L_{\tau}-1$, so
that the original square lattice is replaced by a square prism with
dimensions $L\times L\times L_\tau$. Our Monte Carlo code is able to manage
both periodic and antiperiodic boundary conditions along the
Euclidean-time dimension $\tau$ (instead, boundary conditions along
the space dimensions will always be periodic). The classic spins
interact with an effective coupling $k>0$ that directly relates to the
transverse field through the monotonically decreasing transformation
$$\Gamma=-\frac{1}{2k} \log \mathrm{tanh}(k)\,.$$
Specifically, the probability of a configuration $\boldsymbol{S}$ ($\boldsymbol{S}$ is a shorthand for the $L^2\times L_\tau$ spins in the system)  is given by
\begin{eqnarray}\label{eq:P-S-tau}
  p(\boldsymbol{S})& =& \frac{\text{e}^{-k \mathcal{E}(\boldsymbol{S})}}{Z}\,,\quad Z=\sum_{\lbrace\boldsymbol{S}\rbrace} \text{e}^{-k\mathcal{E}(\boldsymbol{S})}\,,\\\label{eq:S-tau}
\mathcal{E}(\boldsymbol{S}) &=&  - \sum_{\tau=0}^{L_{\tau-1}}\Big[\sum_{\langle \boldsymbol{x},\boldsymbol{y}\rangle}\,
J_{\boldsymbol{x},\boldsymbol{y}} S_{\boldsymbol{x},\tau}S_{\boldsymbol{y},\tau} + \sum_{\boldsymbol{x}} S_{\boldsymbol{x},\tau} S_{\boldsymbol{x},\tau+1}\Big]\,.
\end{eqnarray}
The interested reader will find a more paused exposition of the physical problem in Ref.~\cite{bernaschi:23}.

The numerical study of the system \eqref{IQSG} represents a formidable
challenge, despite of its illusory simplicity, because it is necessary
either to secure an accurate statistics, based on many large scale (in
particular along the imaginary time dimension) Montecarlo simulations
or, in alternative, to find the first eigenvalues of the {\em
  transfer} matrix of the system taking into account that the matrix size grows as $2^{L\times L}$.

Since we did not find suitable existing solutions for working with neither of the two approaches, we
decided to develop from scratch high performance codes both for our
study and for providing them to the community of numerical statistical mechanics.

The rest of the paper is organized as follows: in Section
\ref{sec:qisg} we describe the main codes of the QISG suite; Section
\ref{sec:requir} reports practical information about the building and
the usage of the codes described in Section \ref{sec:qisg}; Section
\ref{sec:perf} contains some sample results and performance data and,
finally, Section \ref{sec:conc} concludes the work also indicating
some directions for further developments. Additional technical details
and extended discussions can be found in the appendices.

\section{The QISG suite \label{sec:qisg}}

We provide two multi GPU codes for the study of a
Quantum Ising Spin Glass in 2D. The main reason for having {\em two} codes
is that the transfer matrix method is {\em exact} but unfeasible for
any $L>6$ (let us recall that $L$ is the linear size of the spin
system).  However, to avoid ``finite size'' effects it is necessary to
study the critical properties of systems with $L>6$. So, we employed the
transfer matrix method to {\em double check} with a two-step procedure
the Montecarlo program. First of all we compared the results obtained
with the transfer matrix approach with those produced by a CPU version
of the Montecarlo program. This has been possible for lattices up to
$L=6$. Then, we compared the results of the CPU Montecarlo program
with those produced by the multi GPU version for larger lattices (our
GPU implementation requires $L$ being an exact multiplier of $4$, so
it was not possible to compare directly $L=6$).  It is useful to
highlight that, due to the features of the computation, the comparison
between the CPU and the GPU can be {\em exact}. The other reason to
develop and use a code for the transfer matrix diagonalization was to
get hints about what is worth to be investigated on larger lattices
where only the Montecarlo method can be employed.  To ease this
comparison, we have diagonalized the version of the transfer matrix
that corresponds to our Trotter-Suzuki scheme~\cite{bernaschi:23} (see
\ref{App:test} for more on this comparison).

\subsection{The MCQSG code \label{sec:MCQSG}}

\begin{figure}
\begin{center}
\begin{tikzpicture}
    \node[rectangle,draw,fill=gray,text=white,minimum width=10cm,minimum height=0.7cm] (c) at (0.15,-0.5) {MPI (for managing multiple disorders)};
    \node[rectangle,draw,fill=RoyalBlue,text=white,minimum width=3cm] (l) at (-3,-2.0) {Task 0 (parallel tempering)};
    \node[text width=1cm] at (0.18,-2.0) {$\ldots$};
    \node[rectangle,draw,fill=RoyalBlue,text=white,minimum width=3cm] (l) at (3.0,-2.0) {Task N-1 (parallel tempering)};
    \draw[thick] (-5.22,-2.34) -- (-5.3,-3);
    \draw[thick] (-0.8,-2.35) -- (0.1,-3);    
    \node[rectangle, text=white,text width=2.6cm, minimum height=2.2cm, pattern=flexcheckerboard_redblue] (l) at (-4.1,-4.15) {{\bf GPU 0\\checkerboard}};
    \node[text width=1cm] at (-2.1,-4.2) {$\ldots$};
    \node[rectangle, text=white,text width=2.6cm, minimum height=2.2cm, pattern=flexcheckerboard_redblue] (l) at (-0.6,-4.15) {{\bf GPU 3\\checkerboard}};
    \node[rectangle,fill=red,draw,text width=3.02cm](l) at (-3.59,-6.0) {multispin coding};
    \node[rectangle,fill=red,draw] (l) at (-5,-6.5) {0};
    \node[rectangle,fill=red,draw] (l) at (-5,-7) {$\downarrow$};    
    \node[rectangle,fill=red,draw] (r) at (-4.6,-6.5) {1};
    \node[rectangle,fill=red,draw] (r) at (-4.6,-7) {$\uparrow$};        
    \node[rectangle,fill=red,draw] (r) at (-4.2,-6.5) {1};
    \node[rectangle,fill=red,draw] (r) at (-4.2,-7) {$\uparrow$};        
    \node[rectangle,fill=red,draw] (r) at (-5+1.225,-6.5) {0};
    \node[rectangle,fill=red,draw] (r) at (-5+1.225,-7) {$\downarrow$};    
    \node[rectangle,fill=red,draw] (r) at (-5+1.635,-6.5) {0};
    \node[rectangle,fill=red,draw] (r) at (-5+1.635,-7) {$\downarrow$};    
    \node[rectangle,fill=red,draw] (r) at (-5+2.035,-6.5) {1};
    \node[rectangle,fill=red,draw] (r) at (-5+2.035,-7) {$\uparrow$};    
    \node[rectangle,fill=red,draw] (r) at (-5+2.435,-6.5) {0};
    \node[rectangle,fill=red,draw] (r) at (-5+2.435,-7) {$\downarrow$};    
    \node[rectangle,fill=red,draw] (r) at (-5+2.815,-6.5) {1};
    \node[rectangle,fill=red,draw] (r) at (-5+2.815,-7) {$\uparrow$};

    \node[rectangle,text=white,fill=blue,draw,text width=3.03cm](l) at (3.5-3.58,-6.0) {multispin coding};
    \node[rectangle,text=white,fill=blue,draw] (l) at (3.5-5,-6.5) {1};
    \node[rectangle,text=white,fill=blue,draw] (l) at (3.5-5,-7) {$\uparrow$};    
    \node[rectangle,text=white,fill=blue,draw] (r) at (3.5-4.6,-6.5) {1};
    \node[rectangle,text=white,fill=blue,draw] (r) at (3.5-4.6,-7) {$\uparrow$};        
    \node[rectangle,text=white,fill=blue,draw] (r) at (3.5-4.2,-6.5) {0};
    \node[rectangle,text=white,fill=blue,draw] (r) at (3.5-4.2,-7) {$\downarrow$};        
    \node[rectangle,text=white,fill=blue,draw] (r) at (3.5-5+1.225,-6.5) {0};
    \node[rectangle,text=white,fill=blue,draw] (r) at (3.5-5+1.225,-7) {$\downarrow$};    
    \node[rectangle,text=white,fill=blue,draw] (r) at (3.5-5+1.635,-6.5) {0};
    \node[rectangle,text=white,fill=blue,draw] (r) at (3.5-5+1.635,-7) {$\downarrow$};    
    \node[rectangle,text=white,fill=blue,draw] (r) at (3.5-5+2.435,-6.5) {1};
    \node[rectangle,text=white,fill=blue,draw] (r) at (3.5-5+2.435,-7) {$\uparrow$};    
    \node[rectangle,text=white,fill=blue,draw] (r) at (3.5-5+2.035,-6.5) {0};
    \node[rectangle,text=white,fill=blue,draw] (r) at (3.5-5+2.035,-7) {$\downarrow$};    
    \node[rectangle,text=white,fill=blue,draw] (r) at (3.5-5+2.835,-6.5) {1};
    \node[rectangle,text=white,fill=blue,draw] (r) at (3.5-5+2.835,-7) {$\uparrow$};    

\end{tikzpicture}
\end{center}
\caption{Multilevel parallelism in MCQSG}\label{fig:mcqsg}
\end{figure}
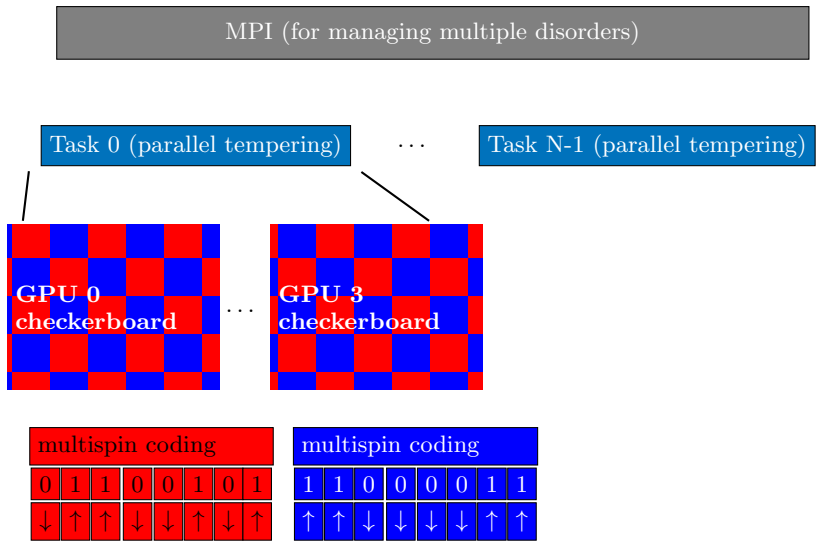

MCQSG is our code for Quantum Ising Spin Glass based on the classic Metropolis algorithm. 
It exploits four levels of parallelism:
\be
\item multi-spin coding;
\item multi (CUDA) threads;
\item multi-GPU;
\item multi-task.
\ee
as shown in Figure \ref{fig:mcqsg}.
Multi-spin coding is not a novel technique, but it remains a very useful {\em
  trick} to speed up simulations of spin systems \cite{JR} (see
Ref.~\cite{preis:09} for a GPU code that follows a diffent approach for spin system simulations).
Since we aim at simulating the Ising model, a single bit is enough to represent a spin, so that a 32 bit word contains
32 spins. Recent work employing multi-spin coding to achieve
  the parallel simulation of a single physical system  includes
  studies of surface-growth models~\cite{pagnani:13,kelling:16} and
  classical spin glasses~\cite{fernandez:15}.
As in other multi-spin coding implementations, the Metropolis factor $e^{-k(\mathcal{E}^\prime-\mathcal{E})}$
is precomputed in our code for all possible values. We have implemented the
Metropolis algorithm as explained in Ref.~\cite{ito:90}, due to the nicely
small number of boolean operations required by that elegant solution. Given
that we are simulating a single physical system,
special attention is reserved to the generation of the random numbers as described in Section \ref{sec:randnumb}.

For MCQSG we resort to a custom memory addressing scheme that we
adapted from a previous one that we proposed in the past for other
spin glass models \cite{multispin}. It maximizes memory coalescing
and eliminates completely costly operations to manage periodic
boundary conditions that might cause threads divergence.

The spins are split in two subsets ({\em red} spins and {\em blue}
spins) according to the classic {\em checkerboard} decomposition that,
by exploiting the fact that each spin interacts only with its nearest
neighbour spins (6 in our case), guarantees that spins updated in
parallel are statistically independent from each other (a necessary condition for the
correctness of the algorithm).
By design, MCQSG works only if the size of the system fulfills some
requirements. In particular,
the imaginary time dimension $L_\tau$ must be
$$ L_\tau=2048*n_t~(n_t~\text{a positive integer})\,.$$
As a matter of fact, $n_t$ rules the number of threads that we may include in a CUDA block,
because to simulate a single spatial site $\boldsymbol{x}$ we need $n_t$ {\em warps}, or $n_t*32$ threads.
Then, we need to write the number of spatial sites as the product of two integers
$$L^2={\tt n\_blocks\_per\_k*n\_sites\_per\_block}$$
so  that the number of threads per block will be
$${\tt num\_threads\_per\_block}=n_t*32*{\tt n\_sites\_per\_block}\,.$$
To be compliant with CUDA best practices, {\em i.e.,} having a number of threads {\em per} block equal to a power of two, we see that both
$n_t$ and ${\tt n\_sites\_per\_block}$ must be powers of two as well. 
For instance, for $L_\tau=2048$ (our choice in most of our production
runs~\cite{bernaschi:23}) we have $n_t=1$. So, for $L=20$, we could
have up to 512 threads per block (with ${\tt n\_sites\_per\_block}=16$
and ${\tt n\_blocks\_per\_k}=25$), or 1024 for $L=16,24$ (in which
case we could set ${\tt n\_sites\_per\_block}=32$ and
${\tt n\_blocks\_per\_k}=8$ or $18$, respectively).\\
It is apparent that, in order to have at least 512 threads per block with
$L_\tau=2048$, we are limited to $L$ being a multiple of four (to have 1024
threads per block $L$ should be a multiple of eight).\\
The number of blocks that will run inside every GPU will be
$${\tt n\_blocks\_per\_k}*\frac{N_k}{4}$$
which ensures that the device {\em occupancy} will be reasonably high (the meaning of the factor $N_k/4$ is explained below).\\
Summarizing, the domain is updated by a total number of CUDA threads equal to
$$\frac{1}{2} \times \frac{1}{32} \times L \times L \times L_\tau\times \frac{1}{4} \times N_{k}$$
The factor $\frac{1}{2}$ comes from
the splitting of the spin in the {\em red} and {\em blue} subsets,
the factor $\frac{1}{32}$ from the employment of multispin coding.
The last two factors (namely $\frac{1}{4}$ and $N_{k}$) come from the application of
the Parallel Tempering (PT) technique to speed-up the thermalization
process \cite{hukushima:96}.  $N_{k}$ is the number of transverse
fields used for the PT. We recall that PT works with several
independent copies of a system. Each system has a different $k$ value selected within a predefined set
[$k$ codes a value of the transverse field in the present case,
see Eqs.~(\ref{eq:P-S-tau}, \ref{eq:S-tau})]. The swap of the system
copy $a$, evolving under coupling $k_a$, with sample $b$ ---that
evolves under coupling $k_b$--- takes place with
probability\footnote{The energy kernel in the MCQSG code actually
  delivers $-\mathcal{E}$, rather than $\mathcal{E}$, hence the additional minus sign that
  the reader will find in the acceptance probability for the
  Parallel-Tempering exchange probability.}
$$
P=\text{min}(1,e^{(k_a-k_b)(\mathcal{E}_a-\mathcal{E}_b)})\,.
$$
Actually, there is no need to swap the spin systems, since it is
sufficient to swap the $k$ values.  In general, we split the
values of the transverse field (usually in the range of a few tens)
in 4 subsets each running on a distinct GPU. This is the origin of the
factor $\frac{1}{4}$.  As a matter of fact, using more than one GPU to
implement the PT technique represents our third level of parallelism.
However, the MCQSG code may run on a single GPU as well. The number of
GPU available for the PT is defined in the main input file described in the \ref{app:inputfile}.

The fourth level of parallelism supports the concurrent execution of
distinct instances of the disorder or multiple instances ({\em
replica}) of the same disorder. We resort to a MPI {\em wrapper} that
executes a copy of MCQSG for each task using different input data. This
level can be considered {\em embarrassingly} parallel. However, it
simplifies the management of the simulations (the user does not need
to write {\em scripts} to automate the execution of multiple runs). The
MPI wrapper can be easily adapted to execute other applications having similar
features.

MCQSG resorts to CUDA streams and asynchronous memory copies
to support overlap between computation, communication and I/O operations.

Besides the above mentioned input file, MCQSG needs two more
files: $i)$ a file which contains the list of values of the transverse field $k$
(used for the PT) and $ii)$ a file which contains the lookup-table required by
the random number generator (see \ref{sec:randnumb}). Information
on the creation of this lookup-table is contained in~\ref{app:LUT}.

The MCQSG code writes periodically the spin configuration that is used
for both off-line analysis and for check-pointing purposes.
The amount of the output data, depends, obviously, on the size of the
spin system ($L \times L \times L_z$).  Each configuration is stored
in a file of a few Mbytes but since a typical run saves 500
configurations, each run requires, at least, a few Gbytes of disk
space. Note that, due to the nature of the spin data, compression does
not help and, as a matter of fact, it would be just a waste of time.

The writing of a configuration is asynchronous with respect to the Monte
Carlo simulation (a separate {\em pthread} is in charge of it).

\subsubsection{Random Numbers \label{sec:randnumb}}

In the original \emph{daemons} multi-spin coding method~\cite{ito:90}
that we have adapted (and modified), the thermal bath is represented
by the two daemon-bits $d_0$ and $d_1$ that form a two-bits integer
$s=d_0+2*d_1$. Hence, Ito and Kanada drew a random number $R$ uniformly 
distributed in $[0,1)$ that is used to get $s$ in the following way:
\begin{align*}
s=0 & \text{ if } R>\text{e}^{-4k} \\
s=1 & \text{ if } \text{e}^{-4k}> R>\text{e}^{-8k} \\
s=2 & \text{ if } \text{e}^{-8k}> R>\text{e}^{-12k} \\
s=3 & \text{ if } \text{e}^{-12k}>R
\end{align*}

In Ito and Kanada's original proposal, the 32 bits in a memory word represent
spins in different samples [{\em i.e.,} 32 different realizations of the
couplings $J_{\boldsymbol{x},\boldsymbol{y}}$ in Eq.~\eqref{IQSG};
this is why this simpler version of the algorithm is some times named
multi-sample multi-spin coding]. The algorithm includes also
two 32-bits daemon words
${\tt id0}$ and ${\tt id1}$. The $j$-th bit in ${\tt id0}$ provides the least
significant bit of a two-bit integer ${\tt s[j]}$ (of course, the
most significant bit of ${\tt s[j]}$ is the $j$-th bit in
${\tt id1}$).  In a multi-sample setting
~\cite{ito:90}, it was a perfectly reasonable choice to draw just two
daemon bits $d_0$ and $d_1$ as explained above, and then set
the 32 bits in ${\tt id0}$ identical to $d_0$ (${\tt id1}$ is formed
in an analogous way from the single bit $d_1$).

However, our MCQSG program leverages multispin coding by packing
32 spins \emph{from the same system} in a 32-bit memory word. In other
words, ours is a multi-site multi-spin coding method. Hence, it is
mandatory (for us) to code 32 \emph{statistically independent}
integers ${\tt s[j]}$ in our daemon words ${\tt id0}$ and ${\tt
  id1}$. A simple solution is just drawing 32
independent random numbers $R$, so that an independent integer $s$ is
obtained from each $R$ but we found out a more efficient approach.

Our solution requires obtaining three statistically independent random
words {\tt b4}, {\tt b8} and {\tt b12}. Our requirements for the 32
bits in these memory words are: \bi
\item all bits should be statistically independent;
\item (each bit) equal to zero with probability equal to $1-e^{-4k}$ \ei The
  procedure we follow to obtain the daemon words {\tt id1} and {\tt id2} from
  {\tt b4}, {\tt b8} and {\tt b12} is explained in~\ref{app:daemon-bits}.
  Thus, our problem boils down to obtaining three 32-bit words {\tt b4}, {\tt
    b8} and {\tt b12} with the prescribed statistical properties. We obtain
  these words through the following procedure.  To obtain the 32 bits in (say)
  {\tt b4}, we use four ``50/50'' 64 bits random words ({\em i.e,} each bit
  has the same probability of being either equal to 0 or equal to 1) obtained
  from our own CUDA implementation of D.E. Shaws's counter-based
  philox\_4x32\_10 parallel generator
  \cite{philox}.\footnote{Our counters correlate with the
      lattice updates (we have a separate counter for every clone in the
      parallel tempering), so that the code provides reproducibility across
      runs with different work-distribution.} We combine
    the 128 bits from philox\_4x32\_10 with another set of 128 bits produced
    by a single iteration of the xoroshiro128++ transformation
    \cite{xoroshiro128++}.\footnote{Specifically, the 32
      most-significant bits of each of our four ``50/50'' 64-bits random words
      are taken from the 128 bits generated by a single philox\_4x32\_10
      iteration. Next, we transform the philox\_4x32\_10 bits through an
      iteration of the xoroshiro128++ transformation (which requires fewer
      operations with respect to the philox\_4x32\_10). The resulting 128 bits
      provide the 32 least-significant bits of each of our four ``50/50''
      64-bits random words.} Then, 8 bit random integers are generated as
  follows: \be
\item To each integer $n=0,1,\ldots,255$
is assigned a probability equal to
$$
P[n]= \text{e}^{-4k*{\text{pop\_count}(n)}} * (1 -\text{e}^{-4k})^{(8-\text{pop\_count}(n))}
$$
where $\text{pop\_count}(n)$ is the {\em population count} function that
returns the number of bits equal to 1 in $n$;\footnote{Hence, every one of the
  8 bits in $n$ is statistically independent, and set to 1 with probability
  $e^{-4k}$. Even for the largest $k$ value in
  our simulations~\cite{bernaschi:23}, $k=0.32$, one has
  $\text{e}^{-4*k*8}=\text{e}^{-10.24}\approx 2^{-14.773}$ which is
  perfectly representable with 64-bit arithmetics.}
\item for each $n$, we build a ``normalized cumulative'' 
$$
\text{cum\_float}(0)= P[0]*2^{64}
$$
$$
\vdots
$$
$$
\text{cum\_float}(n)= \text{cum\_float}(n-1) + 2^{64}*P(n)\,.
$$
However, we need to compare this cumulative with our integer-valued random
numbers. Hence, we compute
$\text{cumulative}(n)$ as the 64-bit  integer truncation of
$\text{cum\_float}(n)$, adding an additional $n=-1$ element to the list that
will be handy for the comparison in the next step:
$$
\text{cumulative}(-1)=-1, \ldots, \text{cumulative}(255)=2^{64}-1;
$$
\item given one of the 64 bit ``50/50'' random words, we find $0\le n^* \le 255$ 
such that\footnote{We set as well $n^*=255$ if the ``50/50'' 64-bit random
word equals $2^{64}-1$.}
$$
\text{cumulative}(n^* -1) \leq \text{64 bit}~ \text{``50/50''}~\text{random word} < \text{cumulative}(n^*)
$$
to speedup the search for $n^*$, we resort to a binary-search combined
with the look-up table that is one of the MCQSG input files.
\ee This
procedure is repeated four times. In the end, we combine the four
$n^*$ in a 32 bit word.

We performed two separate tests of this procedure. Our first (and simple
test), just checked that the statistical properties of the integer
${\tt s[j]}$ were as expected. The second check consisted on the quantitative
comparison of final outcomes (i.e. the quantities we considered in the
statistical analysis of our production runs~\cite{bernaschi:23}) as computed
in two different ways for the very same coupling matrix
$J_{\boldsymbol{x},\boldsymbol{y}}$. One of these computations employed a standard CPU
program. On our second computation, the same quantities were obtained from the MCQSG program
(that implements the above explained procedure). This comparison is discussed
in~\ref{App:test}.

\subsection{The EDQSG code}

The EDQSG code finds the first (four, by default) eigenvalues and
the corresponding eigenvectors for the transfer matrix appropriate
to our Trotter-Suzuki approximation~\eqref{eq:S-tau} (${\cal N}$ is an
irrelevant, but convenient normalization):
\begin{eqnarray}\label{eq:transfer-matrix}
  T&=&\text{e}^{-\frac{k}{2}H_0}\frac{\text{e}^{-k H_1}}{{\cal N}^{L^2}}\text{e}^{-\frac{k}{2}H_0}\,,\quad {\cal N}=2\,\text{e}^{-k}\text{cosh}\, k\,\,\text{cosh}\, k\varGamma\,,\\\nonumber
  H_0&=&-\sum_{\langle \boldsymbol{x}, \boldsymbol{y}\rangle}J_{\boldsymbol{x},\boldsymbol{y}}\sigma^Z_{\boldsymbol{x}}\sigma^Z_{\boldsymbol{y}}\,,\quad H_1=-\varGamma\sum_{\boldsymbol{x}}\sigma^X_{\boldsymbol{x}}\,.
\end{eqnarray}

Since we are interested only in the largest eigenvalues of matrix $T$,
it is sufficient to resort to the classic Lanczos method, an iterative
algorithm for computing the extremal eigenvalues and corresponding
eigenvectors of a large, symmetric matrix
\cite{Lanczos}. What makes the problem very demanding, from the
computational viewpoint, is the matrix size.  We recall that the size
of the transfer matrix is equal to $2^{L\times L}$ where L is the
linear size of the spin system. For L=5 the transfer matrix is a
$2^{25} \times 2^{25}$ matrix and the Lanczos algorithm can be
executed even on a (high end) laptop. Unfortunately, simply moving
from $L=5$ to $L=6$ changes the scenario dramatically, because the
matrix size grows to $2^{36} \times 2^{36}$ ($\sim 69$ billion
rows). Actually, by exploiting the parity symmetry of the transfer
matrix, it is possible to express it as the direct sum of two matrices
having a size that is half of the original matrix. These {\em even}
and {\em odd} matrices can be diagonalized separately. Specifically,
our code obtains the two leading eigenvalues of each sector. We
also obtain the eigenvector corresponding to the leading eigenvalue
within each of the two sectors. Details on the even and odd submatrices (and about our notational conventions) can be found in the~\ref{app:transfer}.

Despite of the splitting in two matrices, the problem remains very
challenging because GPUs have a relatively limited amount of memory
compared to classic CPUs that may have Terabytes of RAM.  However,
there is no need to store the whole matrix since we can follow a
matrix-free approach in which the algorithm for solving a linear system of
equations or an eigenvalue problem does not need to store the coefficient
matrix explicitly, but accesses the matrix by evaluating matrix-vector
products that are the workhorse of the Lanczos's
algorithm. We based our code on a combination of two widely used open
source software packages Petsc \cite{PETSC} and Slepc \cite{SLEPC}.
Our original contribution is a highly-tuned multi-GPU-CPU solution to
compute the (transfer) matrix-vector product. The product is carried
out in three phases: \be
\item {\em Initialization:} the Slepc library (running on CPU) provides an input vector \verb|in[]|,
that is copied to the GPU, Then, on GPU, we execute
\begin{verbatim}
        for(i=0; i<Nloc; i++) {
                psi[i]=in[i]*w[i]; 
        }
\end{verbatim}
where
\bi
\item \verb|Nloc|=$\frac{2^{(L \times L)-1}}{P}$; $L$ is the linear size of the spin system ({\em e.g.,} 5 or 6) and \verb|P| is the number of GPUs;
\item \verb|w[]| is the vector that contains the (precomputed) Boltzmann factors $e^{-k \times H_0/2}$, see Eq.~\eqref{eq:transfer-matrix}, for all possible configurations of spins;
\item \verb|psi[]| is a temporary vector
\ei
\item {\em Computation:} the product is carried out in \verb|L2=|$(L\times L)-1$ {\em rounds}\\
\verb|    for(j=0;j<L2;j++){|\\
\verb|        for(i=0;i<Nloc;i++){|\\
\verb|            scra[i]=jP*psi[i]+jM*psi[|$i\oplus 2^j$\verb|];|\\
\verb|        }|\\
\verb|        swap(scra,psi)|\\
\verb|    }|\\
where
\bi
\item \verb|scra[]| is a temporary vector;
\item \verb|Jp| is equal to $\frac{e^k}{e^k+e^{-k}}$ and \verb|Jm| is equal to $\frac{e^{-k}}{e^k+e^{-k}}$
\item $\oplus$ is the {\em bitwise exclusive OR} (between \verb|i| and $2^j)$ 
\ei
The connection defined by the $i\oplus 2^j$ stride is shown for the first 4 values of $i$ and the first 4 {\em rounds} below

{\scriptsize
\hspace*{0cm}%
\begin{tikzpicture}[
    MyStyle/.style={draw, minimum width=2em, minimum height=2em, 
                outer sep=0pt},
  ]
\begin{scope}[shift={(-12,0)}]
	\node at (0,0.5) {round 0}; \matrix (A) [matrix of math nodes, nodes={MyStyle, anchor=center}, column sep=-\pgflinewidth]
	{0 & 1 & 2 & 3 & 4 & 5 & 6 & 7 & 8 & 9 & 10 & 11 & 12 & 13 & 14 & 15 & \cdots  & N-2 & N-1\\};
\begin{scope}[-]
  \draw[red] (A-1-1.north) [out=25, in=155] to (A-1-2.north);
  \draw[green] (A-1-2.south) [out=-30, in=-155] to (A-1-1.south);
  \draw[blue] (A-1-3.north) [out=35, in=155] to (A-1-4.north);
  \draw(A-1-4.south) [out=-40, in=-155] to (A-1-3.south);
\end{scope}
\end{scope}
\end{tikzpicture}

\hspace*{0cm}\begin{tikzpicture}[
    MyStyle/.style={draw, minimum width=2em, minimum height=2em, 
                outer sep=0pt},
  ]
\begin{scope}[shift={(-12,0)}]  
	\node at (0, 0.5) {round 1}; \matrix (B) [matrix of math nodes, nodes={MyStyle, anchor=center}, column sep=-\pgflinewidth]
	{0 & 1 & 2 & 3 & 4 & 5 & 6 & 7 & 8 & 9 & 10 & 11 & 12 & 13 & 14 & 15 & \cdots  & N-2 & N-1\\};
\begin{scope}[-]
  \draw[red] (B-1-1.north) [out=25, in=155] to (B-1-3.north);
  \draw[green] (B-1-2.south) [out=-30, in=-155] to (B-1-4.south);
  \draw[blue] (B-1-3.north) [out=35, in=155] to (B-1-1.north);
  \draw (B-1-4.south) [out=-40, in=-155] to (B-1-2.south);
\end{scope}
\end{scope}
\end{tikzpicture}

\hspace*{0cm}\begin{tikzpicture}[
    MyStyle/.style={draw, minimum width=2em, minimum height=2em, 
                outer sep=0pt},
  ]
\begin{scope}[shift={(-12,0)}]    
  \node[align=center] at (0, -0.5) {round 2}; \matrix (C) [matrix of math nodes, nodes={MyStyle, anchor=center}, column sep=-\pgflinewidth]
	{0 & 1 & 2 & 3 & 4 & 5 & 6 & 7 & 8 & 9 & 10 & 11 & 12 & 13 & 14 & 15 & \cdots  & N-2 & N-1\\};
\begin{scope}[-]
  \draw[red] (C-1-1.north) [out=25, in=155] to (C-1-5.north);
  \draw[green] (C-1-2.north) [out=30, in=155] to (C-1-6.north);
  \draw[blue] (C-1-3.north) [out=35, in=155] to (C-1-7.north);
  \draw (C-1-4.north) [out=40, in=155] to (C-1-8.north);
\end{scope}
\end{scope}
\end{tikzpicture}

\hspace*{0cm}\begin{tikzpicture}[
    MyStyle/.style={draw, minimum width=2em, minimum height=2em, 
                outer sep=0pt},
  ]
\begin{scope}[shift={(-12,0)}]    
  \node[align=center] at (0, -0.5) {round 3}; \matrix (C) [matrix of math nodes, nodes={MyStyle, anchor=center}, column sep=-\pgflinewidth]
	{0 & 1 & 2 & 3 & 4 & 5 & 6 & 7 & 8 & 9 & 10 & 11 & 12 & 13 & 14 & 15 & \cdots  & N-2 & N-1\\};
\begin{scope}[-]
  \draw[red] (C-1-1.north) [out=25, in=155] to (C-1-9.north);
  \draw[green] (C-1-2.north) [out=30, in=155] to (C-1-10.north);
  \draw[blue] (C-1-3.north) [out=35, in=155] to (C-1-11.north);
  \draw (C-1-4.north) [out=40, in=155] to (C-1-12.north);
\end{scope}
\end{scope}
\end{tikzpicture}

}
Actually, the last {\em round} is slightly different from the previous rounds in the indexing (and with an additional difference between the {\em even} and {\em odd} half of the transfer matrix) as shown in equation \ref{eq:lr}. The interested reader may check directly the code for the details.\\
For small spin systems (up to $L=5$), if the GPU has enough memory, all rounds may be executed on a single GPU ({\em i.e.,} \verb|P=1|). 
However, for $L=6$, the matrix size is such that multiple GPU are required (\verb|P|$>1$). In this case, the first $K=L2-\log_2{P}$ rounds
may run in parallel on each GPU. Then, the temporary vector \verb|psi[]| is copied to the CPU because it
must be exchanged among the different tasks using MPI (the stride $i\oplus 2^j$\ exceeds \verb|Nloc|). It looks like we could copy them back to the GPUs after the exchange but since
{\em each} round after the $K^{th}$ requires  an exchange with a different task, the overhead of copying back and forth the data
from the GPU to the CPU overrides the advantage of running on GPU, so the last rounds, after the $K^{th}$, run in parallel on CPU (using OpenMP).
The value of $K$ (that determines how many rounds are executed on GPU) depends on the memory available on each GPU.
For $L=6$ the computation requires 5 double precision arrays of $2^{35}$ elements: $\sim 1.5$ TB of GPU memory. Even using high-end
GPUs, like the Nvidia A100, at least 20 GPU are required. Obviously, the more GPU are used, the faster is the computation, but, at the same time, the number
of computation rounds that are local to the GPU decreases and so it is necessary to find the best trade off between GPU/CPU efficiency and scalability
that depends also on the network speed (although the MPI scalability is close-to-ideal). With our platform (Infiniband based) we found
that the best tradeoff was reached by using (for $L=6$) 256 GPUs each one having (at least) 32 Gbytes.
\item {\em Finalization:} the final result is computed in parallel on CPU 
   and returned to the invoking SLEPC function in the output vector \verb|out[]| 
\begin{verbatim}
        for(int i=0; i<Nloc; i++) {
            out[i]=psi[i]*w[i];
        }
\end{verbatim}
\ee

Another part of the EDQSG code that runs on GPU is the computation
of the correlation functions. This part requires MPI collective
primitives to carry out the final reduction operations.

The EDQSG code has very limited I/O requirements. The input is limited to $i)$
the seed of the random number generator used to define the set of couplings of
the spin glass (that is, the instance of the disorder) and $ii)$ the value of
the transverse field. The output is also very limited: the value of the first
“even” and “odd” eigenvalues and the values of the correlation functions in
Eq.~\eqref{eq:Q2Q4F_GS}, see~\ref{App:test} for details.

There is no need of restarting the computation since each
diagonalization requires a limited amount of time (about 20 minutes)
on 256 high-end GPUs.

\subsection{Codes for analysis \label{sec:analysis}}

Although the reader might wish to develop his/her own
analysis codes, to easy this task we have decided to include in the
project our own suite of analysis programs.

In fact, we have developed also CUDA codes for off-line analysis of
the configurations generated by MCQSG. These are mono-GPU codes but
highly tuned because the study of the configurations’ overlap is
highly demanding from the computational viewpoint. The quantities
that these programs compute are briefly described in~\ref{App:analysis} (see Ref.~\cite{bernaschi:23} for
a physical discussion).
The appendix contains as well, see Sect.~\ref{App:test}, the discussion of
the test mentioned in Sect.~\ref{sec:randnumb}.

\section{QISG building and usage\label{sec:requir}}

The build environment of both MCQSG and EDQSG is based on simple “classic”
Makefile(s). In both Makefiles the CUDA architecture must be defined using the \verb|CUDA_ARCH| macro.
Depending on the environment, it may be necessary to modify some MACRO ({\em e.g,} \verb|CUDA_DIR|). 

We require, at least, CUDA 11.0, plus Petsc 3.13.3 (compiled for 64 bit indexes) and Slepc 3.13.4 for EDQSG.
MCQSG does not require any other software.

The EDQSG code does not require input files. 
It reads all the required data from the command line.
The following is a possible example
{\footnotesize
\begin{verbatim}
mpirun -n 256 EDQSG -L 6 -Z 0_4_8_16_32_64_128_256_512 -k 0.31\
-s 6009222668047633124 -eps_nev 2
\end{verbatim}
}
where the options define
\bi
\item the linear size of the spin system (\verb|-L|);
\item the value of the parameter $k$, recall Eq.~\eqref{eq:P-S-tau}, related
  to the transverse field  (\verb|-K|);
\item the seed for the initialization of the random couplings (\verb|-s|);
\item the set of Euclidean lengths $L_{\tau}$ (separated by
    \verb|_|) for which an estimate of the correlation functions in Eq.~\eqref{eq:Q2Q4F_GS} is provided;
\item the number of eigenvalues to be computed (passed to Slepc) (\verb|-eps_nev|).
\ei
The order of the options is not relevant. Other options are available (for instance, to dump the eigenvectors).
They are described in a {\em Usage} function.

A typical command line to start the MCQSG program is:
{\footnotesize
\begin{verbatim}
MCQSG 0 1 0 1024 0 k.24x2048 input.24x2048 LUT_for_PRNG_nbits11_NB60.bin
\end{verbatim}
}
where
\bi
\item the first two arguments \verb|0 1| identify the sample and subsample;
\item the third one \verb|0| identifies the {\em replica} (in the $[0...5]$ range);
\item the fourth argument \verb|1024| is the number of threads {\em per} block;
\item the fifth argument \verb|0| may be used to set the CPU threads affinity (see the accompanying README for further details);
\ei
MCQSG requires three input files (sixth, seventh, and eighth
arguments). The contents of
\verb|input.24x2048| are reported in the \ref{app:inputfile}. Details about the other two input
files (containing the set of transverse fields and the look up table for the random number generation) are reported in the MCQSG README included in the archive.

As described in Section \ref{sec:MCQSG}, MPI can be used to start \verb|MCQSG| with different input data read from a file having a line for each MPI task. For instance:
\begin{verbatim}
mpirun -np 6 wrappermpi MCQSG wrapperinput 
\end{verbatim}
starts 6 MPI tasks that execute \verb|MCQSG| with input data read from a \verb|wrapperinput| file having the following
sample structure:
\begin{verbatim}
4 100 0 1024 0 k.24x2048 input.24x2048 LUT_for_PRNG_nbits11_NB60.bin 43100
4 100 1 1024 0 k.24x2048 input.24x2048 LUT_for_PRNG_nbits11_NB60.bin 43100
4 100 2 1024 0 k.24x2048 input.24x2048 LUT_for_PRNG_nbits11_NB60.bin 43100
4 100 3 1024 0 k.24x2048 input.24x2048 LUT_for_PRNG_nbits11_NB60.bin 43100
4 100 4 1024 0 k.24x2048 input.24x2048 LUT_for_PRNG_nbits11_NB60.bin 43100
4 100 5 1024 0 k.24x2048 input.24x2048 LUT_for_PRNG_nbits11_NB60.bin 43100
\end{verbatim}
According to the above description of the \verb|MCQSG| command line, this starts
6 {\em replicas} of the sample 4, subsample 100. The last field in each line of the file represents a \verb|MAXTIME|
value (in seconds). It is used to gracefully stop the execution in batch environments that enforce execution time limits.
In this example, \verb|43100| represents 11 hours and 3500 seconds. MCQSG sets an {\em alarm} to that value. When it
expires, a \verb|SIGTIME| signal is received. The signal handler stops the execution in a clean way preventing any
possible problem coming from an abrupt stop that a batch system could enforce after 12 hours (obviously these values are just
an example).
\section{Sample performance results \label{sec:perf}} 

The most significant performance metrics of the MCQSG code is that it
requires just 0.466 picoseconds/spin-update as running on four A100
GPUs. The normalized time per iteration is 1.864
picoseconds/spin-update, since there is no performance degradation
when 4 GPUs concurrently collaborate in the simulation of the same
system. This performance is obtained for the ideal case of a system with $L=16$ and $L_\tau=8192$.
For a different value of $L$, {\em e.g.,} $24$, (we recall that $L$ must be a multiple of $4$, as explained in Section \ref{sec:MCQSG}) the spin-update time is $0.735$ picoseconds (on the same platform).
As to the EDQSG program, finding the first two eigenvalues of the transfer matrix
for a $6 \times 6$ system (the maximum that can be afforded) takes approximately 800 seconds
by using 256 Nvidia A100. The number of iterations of the Lanczos algorithm is, on average, $\sim 40$.

\section{Conclusions and future perspectives\label{sec:conc}}

We have documented our library of GPU programs for the numerical study of the Quantum 
Spin glass with $S=1/2$, on a square lattice  with binary couplings. Given the
extremely demanding nature of this problem, we have
chosen to work on the equal-coupling Trotter-Suzuki approximating. The two
main programs in our library are EDQSG and MCQSG.

EDQSG is our exact diagonalization program, that obtain the four most dominant
eigenvalues for the Transfer Matrix of our problem. EDQSG delivers as well the
eigenvectors corresponding to the two largest eigenvalues. We employ the
Lanczos algorithm, as implemented in two open source software packages Petsc
and Slepc. Our contribution is a highly-tuned multi GPU implementation of the
``matrix $\times$ vector'' operation. Indeed although the parity symmetry
allows us to halve the size of the matrix, we have demonstrated the use of
EDQSG in problems with transfer matrices of size $2^{35}\times 2^{35}$
($2^{36}\times 2^{36}$ states to the parity pre-conditioning).  Given the huge
quantity of memory needed, a multi-GPU parallelization is mandatory. In fact,
EDQSG runs in parallel in up to 256 NVIDIA cards. Up to the best of our knowledge, the
size of the diagonalized transfer matrices breaks a World record for this kind
of problems.

Our Monte Carlo program is MCQSG. It implements the Metropolis and the
Parallel Tempering algorithm. Although to a lesser degree than our exact
diagonalization problem, multi GPU parallelization is also very effective for
MCQSG. Indeed, running on four NVIDIA A100 cards, MCQSG delivers a
sub-picosecond time per spin-update. The only comparable performance we are
aware of is not obtained on commercial hardware, but on the custom built computer
Janus II~\cite{janus:14}.\footnote{One of the Janus II FPGAs simulates two independent copies
  (i.e. two replicas) of a \emph{ classical} spin glass on cubic lattices of size
  $160\times 160\times 160$,  at a rate of 1.6 picoseconds per spin update.}

Both programs, EDQSG and MCQSG have been instrumental for our numerical
investigation of the quantum spin-glass transition~\cite{bernaschi:23}.

We envisage two possible future developments. First, it would be important to
extend our code to three spatial dimensions where new physical phenomena are
encountered (for instance, the interplay between the thermal and the quantum
spin-glass transition). Second, from a physical point of view, it would be
most convenient to modify MCQSG in such a way that only even-parity states are
sampled. Although this goal requires a substantial modification of the
standard Trotter-Suzuki framework, we think that this parity-restriction
should be possible.

\appendix

\section{MCQSG input file\label{app:inputfile}}
A MCQSG input file contains the following data
\begin{verbatim}
8                       //L system size in the spatial dimension
2048                    //L Euclidean time
10                      //Number of bits used in the LUT
/home/qisg/output/24x2048/ //Path to output data
500                     //Number of configuration to write on disk
10                      //maxit
400                     //mesfr
30                      //ptfr
0                       //flagJ: coupling seed is read (1) or generated (0)
1                       //simulation is a back-up (0), a new run (1), or both (2)
48                      //Number of values of the transverse field
0                       //Seed for generating couplings J
0                       //Seed for generating initial configuration
0                       //Seed for generating MC random numbers
4                       //Number of GPU to run the program
\end{verbatim}
The meaning of many of them is apparent. We report here just a few additional information 
and refer to the README for the details:
\bi
\item a \verb|0| value for the seeds indicates that the seed is computed at run time;
\item The Elementary Monte Carlo Step (EMCS) is composed of \verb|ptfr|
  Metropolis full-lattice sweeps,  followed by a Parallel Tempering update.
  A full-lattice sweep consists of a red-sublattice update, followed by a
  blue-sublattice update.
\item An online measurement of the energy (that we use only to check the run
  sanity) is performed once every \verb|mesfr| consecutive EMCS. We also
  collect at this time, in the array named \verb|history_betas|, the state of
  the Parallel Tempering random walk (i.e.  the current value of $k$ for
  everyone of the $N_k$ system copies in our simulation). The Parallel
  Tempering random walk is used in a stringent equilibration
  test~\cite{fernandez:09,billoire:18}.
\item We average \verb|maxit| consecutive energy measurements, and write this
  average. This is also the time to write the spin configuration and \verb|history_betas|. Hence,
  consecutive configurations in the hard drive are separated by
  \verb|msfr*maxit| EMCS.  \ei

\section{How to obtain the daemon bits}\label{app:daemon-bits}

As explained in Sect.~\ref{sec:randnumb}, we need to obtain 32
independent, and identically distributed integers ${\tt s[j]}$.  We code the 32 ${\tt s[j]}$  in the two 32-bits
daemon words {\tt id1} and {\tt id2}:
${\tt s[j]}={\tt id1}_j+2*{\tt id2}_j$, where ${\tt id1}_j$ and
${\tt id2}_j$ are, respectively, the $j$-th bits of {\tt id1} and {\tt
  id2}. The sought distribution for ${\tt s[j]}$ is
\begin{align*}
  \text{Probability for } {\tt s[j]}=0 \ ({\tt id1}_j=0\,,{\tt id2}_j=0)&:&  1-\text{e}^{-4k} \,,\\
  \text{Probability for } {\tt s[j]}=1 \ ({\tt id1}_j=1\,,{\tt id2}_j=0)&:&  \text{e}^{-4k}-\text{e}^{-8k} \,,\\
  \text{Probability for } {\tt s[j]}=2 \ ({\tt id1}_j=0\,,{\tt id2}_j=1)&:&  \text{e}^{-8k}-\text{e}^{-12k} \,,\\
  \text{Probability for } {\tt s[j]}=3 \ ({\tt id1}_j=1\,,{\tt id2}_j=1)&:&  \text{e}^{-12k} \,.\\
\end{align*}

To achieve our goal, we start from three statistically independent
random-words {\tt b4}, {\tt b8}, {\tt b12} such that: (i) each bit in
the word is statistically independent, and (ii) a given bit is set
to one with probability $\text{e}^{-4k}$. Note that if we manage to
obtain ${\tt id1}_j$ and ${\tt id2}_j$ from the corresponding bits
${\tt b4}_j$, ${\tt b8}_j$ and ${\tt b12}_j$ (by applying boolean
operators to the words {\tt b4}, {\tt b8}, {\tt b12}) the statistical
independence of the 32 {\tt s[j]} will be guaranteed.

Our first step is to apply the boolean AND operator in the following
way (for the sake of shortness we use C language notation):
\begin{verbatim}
  b8=b4&b8;
  b12=b8&b12;
\end{verbatim}
After this manipulation, the new bit ${\tt b8}_j$ will be equal to $1$ only if the two
original bits ${\tt b4}_j$ and ${\tt b8}_j$ were equal to one, which
happens with probability $\text{e}^{-8k}$ (because the two original
bits were statistically independent, implying that their probabilities
to be one get multiplied by the AND operation). Similarly, the new
${\tt b12}_j$ will be one only if the three original bits
${\tt b4}_j$, ${\tt b8}_j$ and ${\tt b12}_j$ were also one, which
happens with probability $\text{e}^{-12k}$. Hence, the probabilities
for the three bits \emph{after} the AND operations are
\begin{align*}
  \text{Probability for }\  {\tt b4}_j=0\,,\ {\tt b8}_j=0\,,\  {\tt b12}_j=0\ &:&  1-\text{e}^{-4k} \,,\\
  \text{Probability for }\  {\tt b4}_j=1\,,\ {\tt b8}_j=0\,,\  {\tt b12}_j=0\   &:&  \text{e}^{-4k}-\text{e}^{-8k} \,,\\
  \text{Probability for }\ {\tt b4}_j=1\,,\ {\tt b8}_j=1\,,\  {\tt b12}_j=0\   &:&  \text{e}^{-8k}-\text{e}^{-12k} \,,\\
  \text{Probability for }\  {\tt b4}_j=1\,,\ {\tt b8}_j=1\,,\  {\tt b12}_j=1\   &:&  \text{e}^{-12k} \,.\\
\end{align*}
The above probabilities exactly match our expectations for the
${\tt s[j]}$. So, we have fullfiled our task if we are able to
devise a sequence of boolean operations such that
\begin{align*}
  {\tt b4}_j=0\,,\ {\tt b8}_j=0\,,\  {\tt b12}_j=0\ &\text{ ensures that } {\tt d1}_j=0\,,\  {\tt d2}_j=0\,,\\
  {\tt b4}_j=1\,,\ {\tt b8}_j=0\,,\  {\tt b12}_j=0\ &\text{ ensures that } {\tt d1}_j=1\,,\  {\tt d2}_j=0\,,\\
  {\tt b4}_j=1\,,\ {\tt b8}_j=1\,,\  {\tt b12}_j=0\ &\text{ ensures that } {\tt d1}_j=0\,,\  {\tt d2}_j=1\,,\\
  {\tt b4}_j=1\,,\ {\tt b8}_j=1\,,\  {\tt b12}_j=1\ &\text{ ensures that } {\tt d1}_j=1\,,\  {\tt d2}_j=1\,.\\
\end{align*}
Building explicitly the truth table shows that the following C
instructions produce exactly that result:
\begin{verbatim}
  id2=b8;
  id1=(b4^b8)|b12; 
\end{verbatim}
where \verb|^| is the {\em bitwise} Exclusive OR and \verb_|_ is the {\em bitwise} OR operator.
\section{The LUT for the generation of random bits}\label{app:LUT}
Our MCQSG program is complemented with a C program that creates the lookup
table necessary for the generation of the random bits (see Sect.~\ref{sec:randnumb}
for an explanation of the terminology used below).

Specifically, we include the program \texttt{create\_LUT.c} that requires the GNU
quadmath library for 128-bit floating point arithmetics, and the
script \\\texttt{compile\_and\_generate\_LUT.sh} for its compilation and
execution. The script should be invoked with three
arguments, namely a number of bits {\tt nbits}, the number of $k$ values in
the parallel tempering simulation (named \texttt{NK}) and
the name of the file that contains the $k$ values.
For instance,\\
\texttt{sh ./compile\_and\_generate\_LUT.sh 11 32 k.8x2048}\\
generates (and executes) the program \texttt{create\_LUT\_nbits11\_NK32}, that
includes in its name the choices {\tt nbits}=11 and {\tt
  NK}=32.\footnote{While the choice of {\tt NK} varies for the
  different lattice sizes, our production runs~\cite{bernaschi:23} were
  carried out with {\tt nbits}=11.}

Our list of $k$-values (in decreasing order)
should occupy the {\tt NK} first lines of an ASCII file (let us name it
{\tt k.8x2048}, for instance). Hence, the program is invoked as
\texttt{./create\_LUT\_nbits11\_NK32 k.8x2048}\\
to create the binary file \texttt{LUT\_for\_PRNG\_nbits11\_NK32.bin}
that contains the lookup table needed by the MCQSG program.

What \texttt{create\_LUT\_nbits11\_NK32}
does  for every one of the $k$ values read from {\tt k.8x2048}
is:
\begin{enumerate}
  \item Compute $\text{cum\_float}(n)$ with 128-bit floating point
    arithmetics.
  \item Truncate $\text{cum\_float}(n)$ to a 64-bit integer word
    $\text{cumulative}(n)$. Mind that the two extremal values $n=-1$ and 255
    are known beforehand, hence the program  carries out the computation only for
    $n=0,1,\ldots,254$. The 64-bit $\text{cumulative(n)}$ is split into its 32
    most-significant bits and its 32 least-significant bits, so that in the GPU
    one can perform comparisons using 32-bit integer types.
  \item Given the {\tt nbits}=11 most significant bits of the ``50/50'' 64-bit
    random word, determine the largest integer $n_{\text{min}}$ and the smallest
    integer $n_{\text{max}}$ such that one is guaranteed to have
    $$n_{\text{min}}\leq n^*\leq n_{\text{max}}\,.$$ Note that this step should be
    carried out $2^{\tt nbits}$ times.
  \item Pack the 255 8-byte integer words $\text{cumulative}(n)$, and the $2^{\tt
      nbits}$ pairs of 1-byte integer words $(n_{\text{min}},n_{\text{max}})$,  into the {\tt
      uint4} CUDA types that we employ in the GPU.
\end{enumerate}
The resulting lookup table is written in a file in binary format. This file
contains the {\tt NK} $k$-values for which the lookup table was generated,
in order to allow for double-checking at production time.

\section{The parity-aware basis and the transfer matrix}\label{app:transfer}

Let us start by introducing some notations that will help us to describe our
study of the Transfer Matrix in Eq.~\eqref{eq:transfer-matrix}.

As an alternative to
the two-dimensional Cartesian coordinates $\boldsymbol{x}=(x_1,x_2)$ we
may use a one-dimensional lexicographic index
\begin{equation}
  j=x_1+L*x_2\,,\quad j=0,1,\ldots L^2-1\ (\text{because } x_1,x_2=0,1,...L-1)\,.
\end{equation}
Note that the transformation $j\leftrightarrow (x_1,x_2)$ is
one-to-one.  Hence, we shall choose the site-labelling  (either lexicographic or Cartesian) that yields the simplest expressions.

We shall code an element of the computational
basis ({\em i.e.,} the basis that diagonalizes the $L^2$ matrices $\sigma_j^Z$)
through a $L^2$-bits integer ${\tt n}$. If the $j$-th bit in ${\tt n}$
is one, then the corresponding eigenvalue for $\sigma_j^Z$ will be
$S^{\tt n}_j=-1$ (or, switching to the Cartesian site-labelling, $S^{\tt n}_{\boldsymbol x}=-1$). Instead, the eigenvalue is $S^{\tt n}_j=+1$ if the $j$-th bit vanishes. The two factors $\text{e}^{-k H_0/2}$ in Eq.~\eqref{eq:transfer-matrix} are diagonal in the computational basis:
\begin{equation}
  \text{e}^{-k H_0/2}|{\tt n}\rangle = \text{e}^{\frac{k}{2}\sum_{\langle\boldsymbol{x},\boldsymbol{y}\rangle} J_{\boldsymbol{x},\boldsymbol{y}} S^{\tt n}_{\boldsymbol x}S^{\tt n}_{\boldsymbol y}}\, |{\tt n}\rangle\,.
\end{equation}
The central factor in Eq.~\eqref{eq:transfer-matrix} takes the form:
\begin{equation}\label{eq:expH1}
  \frac{\text{e}^{-k H_1}}{{\cal N}^{L^2}}=\prod_{j=0}^{L^2-1}\,\Big[ \frac{\text{e}^k}{\text{e}^k+\text{e}^{-k}}\boldsymbol{1}_j\ +\ \frac{\text{e}^{-k}}{\text{e}^k+\text{e}^{-k}}\sigma^X_j\Big]\,,
\end{equation}
where $\boldsymbol{1}_j$ is the $2\times 2$ identity matrix acting on site $j$
[mind that all the matrices in Eq.~\eqref{eq:expH1} are mutually conmuting].
Now, the action of the matrix $\sigma^X_j$ on the computational basis
is more simple to implement in the $C$ programming language since it corresponds to the flip of the $j$-th bit:
\begin{equation}
  \sigma^X_j|{\tt n}\rangle = |{\tt n}\verb|^|(1\verb|<<|j)\rangle
\end{equation}

Unfortunately, the diagonal basis does not fit our needs, because the parity operator,
\begin{equation}
  P= \prod_{j=0}^{L^2-1}\, \sigma^X_j\,,
\end{equation}
is not diagonal in this basis. Yet, parity is a most prominent symmetry for
our problem. Therefore, we need to chose a basis suitable to $P$. In order to
introduce the new basis, let us consider a $(L^2-1)$-bits integer ${\tt n}$
(as a matter of fact, $|{\tt n}\rangle$ can be regarded as an element of the
computational basis, constrained to have $S^{\tt
  n}_{j=L^2-1}=1$). We shall also need two bit-inversion operations:
\begin{equation}
  \overline{{\tt n}}=2^{L^2}-1-{\tt n}\,,\quad \overline{\overline{{\tt n}}}=2^{L^2-1}-1-{\tt n}\,.
\end{equation}
In words, $\overline{{\tt n}}$ regards ${\tt n}$ as a $L^2$-bit integer (with
the most-significant bit set equal to zero) and inverts all the $L^2$ bits. Instead, $\overline{\overline{{\tt n}}}$ regards ${\tt n}$ as a $(L^2-1)$-bits word,
and inverts the value of the $(L^2-1)$ bits. Hence, the new basis for the
even subspaces is
\begin{equation}
  |{\tt n}\rangle_{\text{e}}=\frac{ |{\tt n}\rangle \ +\ |\overline{{\tt n}}\rangle}{\sqrt{2}}\,,\ {\tt n}=0,1,\ldots\,,2^{L^2-1}-1\,,\quad P  |{\tt n}\rangle_{\text{e}}=+|{\tt n}\rangle_{\text{e}}\,,
\end{equation}
whereas, for the odd subspace, we have
\begin{equation}
  |{\tt n}\rangle_{\text{o}}=\frac{ |{\tt n}\rangle \ -\ |\overline{{\tt n}}\rangle}{\sqrt{2}}\,,\ {\tt n}=0,1,\ldots\,, 2^{L^2-1}-1\,,\quad P  |{\tt n}\rangle_{\text{o}}=-|{\tt n}\rangle_{\text{o}}\,.
\end{equation}
A nice feature of the new basis is that almost all the crucial operators
behave exactly as they did in the computational basis (for
$\text{e}^{-k H_0/2}$, think of ${\tt n}$ as a $L^2$-bit integer with
the most-significant bit set to be zero):
\begin{eqnarray}
  \text{e}^{-k H_0/2}|{\tt n}\rangle_{\text{e},\text{o}} &=& \text{e}^{\frac{k}{2}\sum_{\langle\boldsymbol{x},\boldsymbol{y}\rangle} J_{\boldsymbol{x},\boldsymbol{y}} S^{\tt n}_{\boldsymbol x}S^{\tt n}_{\boldsymbol y}}\, |{\tt n}\rangle_{\text{e},\text{o}}\,,\\
  \sigma^X_j|{\tt n}\rangle_{\text{e},\text{o}} &=& |{\tt n}\verb|^|(1\verb|<<|j)\rangle_{\text{e},\text{o}} \quad\text{ if } j<L^2-1\,.
\end{eqnarray}
The only exception is $\sigma^X_{j=L^2-1}$:
\begin{equation}\label{eq:lr}
  \sigma^X_{j=L^2-1}|{\tt n}\rangle_{\text{e}}=|\overline{\overline{{\tt n}}}\rangle_{\text{e}}\,,\quad \sigma^X_{j=L^2-1}|{\tt n}\rangle_{\text{o}}=-|\overline{\overline{{\tt n}}}\rangle_{\text{o}}\,.
\end{equation}

A final word of caution regards the integer types when implementing
these operations. While  ${\tt n}$ is
representable in a 32-bits word for lattice size $L\leq 5$, already for $L=6$ we shall need a
64-bits word (for the same reason, $1\verb|<<|j$ will need to be replaced by
$1{\tt ULL} \verb|<<|j$).

\section{Our analysis programs}\label{App:analysis}

Our suite for the analysis of the configurations generated by the
MCQSG program contains four programs: {\tt matrix}, {\tt
  coverlap\_pt}, {\tt corr\_euclidea}, and {\tt overlap\_tau}.

We define in~\ref{App:definitions} the quantities computed by 
these programs. In~\ref{App:trick} we explain a simple, but effective,
{\em trick} that helped us to speed-up our CUDA-based analysis codes. Finally,
in~\ref{App:test} we employ our analysis suite to test our main
programs, MCQSG and EDQSG.

\subsection{Definition of the computed quantities}\label{App:definitions}

To lighten notations, let us focus on a single $k$ value. Let us assume that
you have read from disk the whole set of already equilibrated
configurations at Monte Carlo times $t=0,1,\ldots
N_{\text{MC}}-1$. So, an equilibration transient has been already
discarded. Of course, consecutive $t$ are separated by a consistent
number of Monte Carlo steps, so that the spin configurations at
consecutive $t$ are (close to be) statistically independent. A typical
number in our case is $N_{\text{MC}}=250$. We have these $N_{\text{MC}}$
configurations for each one of our $N_{\text{R}}=6$ replicas (replicas are configurations
obtained from statistically independent simulations for \emph{the
  same} coupling matrix $J_{\boldsymbol{x},\boldsymbol{y}}$; for
shortness a choice of the coupling matrix will be named a sample). In short,
we assume that we have, at our disposal, the $N_{\text{MC}}\times N_{\text{R}}$
configurations corresponding to our $k$ value: 
$$S_{\boldsymbol{x},\tau}(t,a)\text{ with } t=0,1,...,N_{\text{MC}}-1\,,\ a=0,1,\ldots,N_{\text{R}}-1\,.$$

Thus, our program {\tt matrix} computes the $N_{\text{R}}$ correlation matrices
$M^{(a)}$, of size $L^2\times L^2$:
\begin{equation}
  M^{(a)}_{\boldsymbol{x},\boldsymbol{y}}=\frac{1}{N_{\text{MC}}}\frac{1}{L_\tau}
  \sum_{t=0}^{N_{\text{MC}}-1} \sum_{\tau=0}^{L_\tau-1} S_{\boldsymbol{x},\tau}(t,a)S_{\boldsymbol{y},\tau}(t,a)\,.
\end{equation}

In order to explain what
our program  {\tt coverlap\_pt} computes,  we shall need two intermediate quantities
\begin{eqnarray}
  q^{(a,b)}_{\boldsymbol{x}}(\tau_1,\tau_2;t_1,t_2)&=& S_{\boldsymbol{x},\tau_1}(t_1,a)S_{\boldsymbol{x},\tau_2}(t_2,b) \,,\label{eqA:q+2+2}\\
  Q^{(a,b)}(\tau_1,\tau_2;t_1,t_2)&=&\sum_{\boldsymbol{x}} q^{(a,b)}_{\boldsymbol{x}}(\tau_1,\tau_2;t_1,t_2)\,,
\end{eqnarray}                                      
from which one would ideally compute [with notations ${\cal M}=\frac{N_{\text{R}}(N_{\text{R}}-1)N_{\text{MC}}^2L_\tau^2}{2}$ and $\boldsymbol{x}=(x_1,x_2)$]
\begin{eqnarray}
  Q_2&=&\frac{1}{{\cal M}} \sum_{b>a} \sum_{t_1,t_2=0}^{N_{\text{MC}}-1} \sum_{\tau_1\,\tau_2=0}^{L_\tau-1} [Q^{(a,b)}(\tau_1,\tau_2;t_1,t_2)]^2\,,\\
  Q_4&=&\frac{1}{{\cal M}} \sum_{b>a} \sum_{t_1,t_2=0}^{N_{\text{MC}}-1} \sum_{\tau_1\,\tau_2=0}^{L_\tau-1} [Q^{(a,b)}(\tau_1,\tau_2;t_1,t_2)]^4\,,\\
  F&=& \frac{1}{2{\cal M}} \sum_{b>a} \sum_{t_1,t_2=0}^{N_{\text{MC}}-1} \sum_{\tau_1\,\tau_2=0}^{L_\tau-1} \Big(\Big|\sum_{\boldsymbol{x}} \text{e}^{\text{i}\frac{2\pi x_1}{L}}  q^{(a,b)}_{\boldsymbol{x}}(\tau_1,\tau_2;t_1,t_2)\Big|^2\nonumber\\
  &+& \Big|\sum_{\boldsymbol{x}} \text{e}^{\text{i}\frac{2\pi x_2}{L}}  q^{(a,b)}_{\boldsymbol{x}}(\tau_1,\tau_2;t_1,t_2)\Big|^2\Big)\,.
\end{eqnarray}
The reason why we specify ``ideally'', is that we have used $L_\tau$ as large as $L_\tau=2048$ in
our $L=24$ simulations~\cite{bernaschi:23} (and even larger $L_\tau$ for
$L=16$). It is clear that, for such a large $L_\tau$, the above computations
would be just too costly. Indeed, the number of operations grow quadratically
with both $N_{\text{MC}}$ and $L_\tau$. Of course, the alert reader will note
that $Q_2$ and $F$ can be computed from the correlation matrix $M^{(a)}$ (that
can be computed with a cost linear in $N_{\text{MC}}$ and
$L_\tau$).\footnote{For instance,
  $Q_2=\frac{2}{N_{\text{R}}(N_{\text{R}}-1)}\sum_{b>a} \text Tr
  M^{(a)}M^{(b)}$.} The reason for keeping {\tt coverlap\_pt} in our suite is
that $Q_4$ enters crucially in the computation of the so-called Binder
cumulant. Now, in order to alleviate the burden of the cost that is quadratic in $L_\tau$, we have
restricted the average over $\tau_1$ and $\tau_2$ to a 128$\times$128
planes subset of constant $\tau_1$ and $\tau_2$. The planes are chosen randomly and
(to mitigate correlations) are kept as distant as possible, taking into account
the geometry of our lattice. Of course, we have changed accordingly the
normalization constant as
${\cal M}=\frac{N_{\text{R}}(N_{\text{R}}-1)N_{\text{MC}}^2128^2}{2}$.

Our last two programs employ the following intermediate quantities
\begin{equation}
  \tilde{q}^{(a)}_{\boldsymbol{x}}(\tau_1,\tau;t)=
  S_{\boldsymbol{x},\tau_1}(t,a)S_{\boldsymbol{x},\tau_1+\tau}(t,a)\,,
\end{equation}
where we use the periodic boundary conditions along the $\tau$ axis\footnote{In the case of simulations
  with antiperiodic boundary conditions along the Euclidean time, we have
  considered only products of an even number of $\tilde{q}$ factors, which
  makes inmaterial the difference betwen periodic and antiperiodic boundary
  conditions.} to
give a meaning to the sum $\tau_1+\tau$.

Thus, our program {\tt corr\_euclidea} computes the Euclidean correlation
function for odd operators:
\begin{equation}
  C^{(a)}_{\boldsymbol{x}}(\tau)=\frac{1}{N_{\text{MC}}}\frac{1}{L_\tau}
  \sum_{t=0}^{N_{\text{MC}}-1} \sum_{\tau_1=0}^{L_\tau-1} \tilde{q}^{(a)}_{\boldsymbol{x}}(\tau_1,\tau;t) 
\end{equation}
At the end, one could average over the replica index $a$, as
well. However, we prefer to separate the computation for the different
replicas in order to compute error estimates for a given
sample.

Finally, our program {\tt overlap\_tau} produces a $\tau$-dependent version
of the quantities $Q_2$, $Q_4$ and $F$. Let us use the shorthand
$$\tilde{Q}^{(a)}(\tau_1,\tau;t)=\sum_{\boldsymbol{x}} \tilde{q}^{(a)}_{\boldsymbol{x}}(\tau_1,\tau;t)\,,$$
from which we compute (the superscript $s$ stresses that the computation is
carried out for a single sample $s$)
\begin{eqnarray}\
  Q_2^{(s,a)}(\tau)&=&\frac{1}{L_\tau N_{\text{MC}}} \sum_{t=0}^{N_{\text{MC}}-1}
                       \sum_{\tau_1=0}^{L_\tau-1}\big[\tilde{Q}^{(a)}(\tau_1,\tau;t)\big]^2\,,\\
 Q_4^{(s,a)}(\tau)&=&\frac{1}{L_\tau N_{\text{MC}}} \sum_{t=0}^{N_{\text{MC}}-1}
                      \sum_{\tau_1=0}^{L_\tau-1}\big[\tilde{Q}^{(a)}(\tau_1,\tau;t)\big]^4\,,\\
 F^{(s,a)}(\tau) &=& \frac{1}{2L_\tau N_{\text{MC}}}
                     \sum_{t=0}^{N_{\text{MC}}-1} \sum_{\tau_1=0}^{L_\tau-1} \Big(\Big|\sum_{\boldsymbol{x}}
                     \,\text{e}^{\text{i}\frac{2\pi x_1}{L}}
                     \tilde{q}^{(a)}_{\boldsymbol{x}}(\tau_1,\tau;t)\Big|^2\nonumber\\
                   &+& \Big|\sum_{\boldsymbol{x}}
                     \,\text{e}^{\text{i}\frac{2\pi x_2}{L}}
                     \tilde{q}^{(a)}_{\boldsymbol{x}}(\tau_1,\tau;t)\Big|^2\Big)\,.
\end{eqnarray}
We keep the computation separate for each replica in order to compute
statistical errors for our  final estimates
\begin{eqnarray}
  Q_2^{(s)}(\tau)&=&\frac{1}{N_{\text{R}}}\sum_a\,
                     Q_2^{(s,a)}(\tau)\,,\label{eq:Q2-tau}\\
  F^{(s)}(\tau)&=&\frac{1}{N_{\text{R}}}\sum_a\,
  F^{(s,a)}(\tau)\,,\label{eq:F-tau}\\
  Q_4^{(s)}(\tau)&=&\frac{1}{N_{\text{R}}}\sum_a\,
                     Q_4^{(s,a)}(\tau)\,.\label{eq:Q4-tau}
\end{eqnarray}

\subsection{A trick to speedup the CUDA implementation}\label{App:trick}

In our program {\tt coverlap\_pt}, we decided that a single thread
would take care of all the operations corresponding to a pair of
Euclidean times $(\tau_1,\tau_2)$ and a pair of Monte Carlo times
$(t_1,t_2)$, recall Eq.~\eqref{eqA:q+2+2}. The 1024 threads in a block
share the assignment of Monte Carlo times $(t_1,t_2)$ and build the pairs
$(\tau_1,\tau_2)$ by
exhausting all the possible combinations of a list of 32 possible
values of $\tau_1$ with another list of 32 values for $\tau_2$. In
other words, given a replica index $a$, it suffices to bring to the GPU shared
memory just 64 planes of spins $S_{\boldsymbol{x},\tau}(t,a)$: 32
planes with fixed $t=t_1$ and $\tau=\tau_1$, the other 32 planes with
fixed $t=t_2$ and $\tau=\tau_2$. The information contained in these
64 planes is enough for the 1024 threads to carry out all their
intended operations. A simple variation of the same idea works for
{\tt corr\_euclidea} and {\tt overlap\_tau}.

\subsection{Consistency checks for EDQSG and MCQSG}\label{App:test}

\begin{figure}[h]
\includegraphics[width=0.7\textwidth]{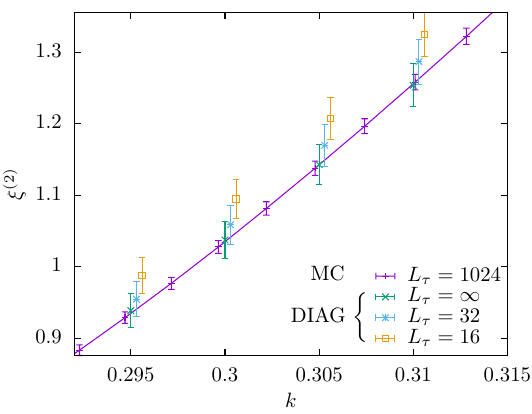}
\caption{Correlation length $\xi^{(2)}$, see Eq.~\eqref{eqA:xi2}, versus $k$
  as computed for $L=6$ lattices, both from Monte Carlo simulation on CPU and from
  exact diagonalization with EDQSG. The estimate from Monte Carlo simulation
  were obtained with Euclidean-time dimension $L_\tau=1024$, by averaging over 1280 samples (the
  continuous line is a cubic-spline interpolation). The results from exact
  diagonalization are an average over 320 samples ($L_\tau=\infty$ results are
  obtained from the eigenvector corresponding to the largest eigenvalue of the
  transfer matrix; the finite $L_\tau$ estimates are obtained by considering
  as well the eigenvector corresponding to the largest eigenvalue of the odd
  submatrix). To ease visualization, the data from exact diagonalization and
  $L_\tau=16$ and 32 were horizontally displaced by a constant amount.
}
\label{fig:check-diag}
\end{figure}

We have carried out two important consistency checks for our main programs
EDQSG and MCQSG.

In the case of EDQSG, we first computed $Q_2$, $Q_4$ and $F$ from a (standard
CPU) Monte Carlo program on a lattice $L=4$ and $L=6$ and $L_\tau=1024$. This $L_\tau$
turns out to be big enough to allow a direct comparison with the outcome of
EDQSG. Indeed, let $|GS\rangle$ be the eigenvector of the transfer matrix
corresponding to its largest eigenvalue, and let
\begin{eqnarray}
  M_{\boldsymbol{x},\boldsymbol{y}}&=&\langle GS| \sigma^Z_{\boldsymbol{x}}
                                       \sigma^Z_{\boldsymbol{y}}|GS\rangle\,,\\
  \tilde{M}_{\boldsymbol{x},\boldsymbol{y}}&=&\frac{1}{2}M_{\boldsymbol{x},\boldsymbol{y}}\Big[\cos\Big(\frac{2\pi}{L}(x_1-y_1)\Big)+\cos\Big(\frac{2\pi}{L}(x_2-y_2)\Big)\Big]\,,\\[1mm]
  R_{\boldsymbol{x},\boldsymbol{y},\boldsymbol{z},\boldsymbol{u}}&=&\langle GS| \sigma^Z_{\boldsymbol{x}}
                                       \sigma^Z_{\boldsymbol{y}}
                                        \sigma^Z_{\boldsymbol{z}} \sigma^Z_{\boldsymbol{u}}|GS\rangle\,,
\end{eqnarray}
Then, in the limit that concern us~\cite{bernaschi:23}, namely
$L_\tau\to\infty$, one has
\begin{equation}\label{eq:Q2Q4F_GS}
  Q_2=\text{Tr} M^2\,,\quad F=\text{Tr} M\tilde{M}\,,\quad
  Q_4=\sum_{\boldsymbol{x},\boldsymbol{y}\boldsymbol{z},\boldsymbol{u}}
  R^2_{\boldsymbol{x},\boldsymbol{y}\boldsymbol{z},\boldsymbol{u}}\,.
\end{equation}
From them, one may compute the finite-lattice correlation length (see, e.g.,
Ref.~\cite{amit:05})
\begin{equation}\label{eqA:xi2}
  \xi^{(2)}=\frac{1}{2\sin\frac{\pi}{L}}\sqrt{\frac{\overline{Q_2}}{\overline{F}}-1}\,
\end{equation}
that we show in Figure \ref{fig:check-diag} (the overlines indicate averages
over samples). The figure carries two important messages. First, it shows that
the limit $L_\tau\to\infty$ can be certainly reached within our numerical
accuracy. And second, the Monte Carlo and Exact Diagonalization results are
statistically compatible. Note that the results in Fig.~\ref{fig:check-diag}
were obtained after averaging over samples. We have strengthened our test by
comparing $Q_2$, $Q_4$ and $F$, as computed in the same sample with both
approaches. Also at this finer level, statistical compatibility was found (at
the $0.01\%$ level of accuracy reached in our Monte Carlo simulations).

\begin{figure}[t]
\includegraphics[width=\textwidth]{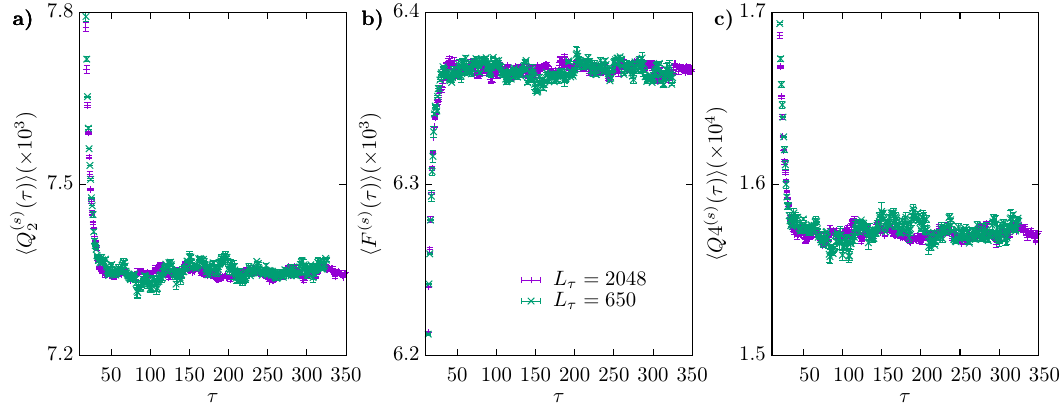}
\caption{The time-correlation functions defined in Eqs.~\eqref{eq:Q2-tau} ({\bf a}),
  ~\eqref{eq:F-tau} ({\bf b}) and ~\eqref{eq:Q4-tau} ({\bf c})
  versus $\tau$, as computed
  \emph{for the same sample} from a standard CPU code $(L_\tau=650)$ and from
  the MCQSG program $(L_\tau=2048)$ at $k=0.29$ which is, approximately, the
  critical point for the model in Eq.~\eqref{IQSG}, see Ref.~\cite{bernaschi:23}. 
}
\label{fig:check-MC}
\end{figure}

As for the check of MCQSG, we have compared the time-correlation functions defined in Eqs.~\eqref{eq:Q2-tau},
  ~\eqref{eq:F-tau} and ~\eqref{eq:Q4-tau} ({\bf c}) with a standard
  CPU program on a lattice with $L=20$ and $L_\tau=650$. The same
  computation, on the same sample,
  has been carried out from the configurations produced by  MCQSG on a lattice
  with $L=20$ and $L_\tau=2048$. The two computations are compared in
  Fig.~\ref{fig:check-MC}. The statistical compatibility of the results at the
  plateaux, as well as the fact that the curves from the two programs reach
  their plateaux at the same $\tau$, is a most reasurring (and significant) check of overall consistency.

\begin{acknowledgments}
  We benefited from two EuroHPC computing grants: specifically, we had access
  to the Meluxina-GPU cluster through grant EHPC-REG-2022R03-182 (158306.5 GPU
  computing hours) and to the Leonardo facility (CINECA) through a LEAP ({\em Leonardo Early Access Program}) grant.
  Besides, we received a small grant
  (10000 GPU hours) from the \emph{Red Española de Supercomputación}, through
  contract no.~FI-2022-2-0007. Finally we thank Gianpaolo Marra for the access
  to the Dariah cluster in Lecce.

  This work was partially supported by Ministerio de Ciencia,
  Innovaci\'on y Universidades (Spain), Agencia Estatal de
  Investigaci\'on (AEI, Spain,\\ 10.13039/501100011033), and European
  Regional Development Fund (ERDF, A way of making Europe) through
  Grant PID2022-136374NB-C21.  This research has also been supported
  by the European Research Council under the European Unions Horizon
  2020 research and innovation program (Grant No. 694925—Lotglassy,
  G. Parisi). IGAP was supported by the Ministerio de Ciencia,
  Innovaci\'on y Universidades (MCIU, Spain) through FPU grant
  No.~FPU18/02665. MB gratefully acknowledges CN1 -- Centro Nazionale
  di Ricerca in High-Performance Computing Big Data and Quantum
  Computing (Italy)-- for support.
\end{acknowledgments}

\end{document}